\documentclass[12pt]{iopart}
\usepackage{epsfig}

\begin{document}

\title[Nanotube Y junctions]{Transport
theory of carbon nanotube Y junctions}

\author{R Egger${}^1$, B Trauzettel${}^2$, S Chen${}^1$ and F Siano${}^1$}
\address{${}^1$ Institut f\"ur Theoretische Physik,
Heinrich-Heine-Universit\"at, D-40225 D\"usseldorf, Germany}
\address{${}^2$ Physikalisches Institut, Albert-Ludwigs-Universit\"at,
D-79104 Freiburg, Germany}

\begin{abstract}
We describe a generalization of Landauer-B\"uttiker theory for
networks of interacting metallic carbon nanotubes.
We start with symmetric starlike junctions and then extend our
approach to asymmetric systems. While the symmetric case is
solved in closed form, 
the asymmetric situation is treated by a mix of perturbative
and non-perturbative methods.   For $N>2$ repulsively interacting
 nanotubes, the only stable fixed point of the symmetric system
corresponds to an isolated node. 
Detailed results for both symmetric and asymmetric systems
are shown for $N=3$, corresponding to carbon nanotube Y junctions.
\end{abstract}
\submitto{\NJP}
\maketitle

\section{Introduction}
\label{intro}

Landauer-B\"uttiker (LB) scattering theory \cite{datta} marks a
 powerful approach to multi-terminal transport in 
conventional (Fermi liquid) mesoscopic devices, but is well-known
to be inapplicable to strongly correlated electron systems. 
Progress along this direction has recently been reported for interacting
 metallic 1D systems, where a good description in terms of a
Luttinger liquid (LL) \cite{gogol} is often available. 
Single-walled carbon nanotubes (SWNTs) represent 
a prime example for such a system \cite{dekker,egger97,tube}.
For a two-terminal setup, the generalization of the 
Landauer formalism to a LL (containing an impurity)
 coupled to noninteracting leads has been
formulated and solved previously \cite{egger96,egger00}. 
For related work, see
 Refs.~\cite{safi,maslov,pono,furu,hekk,oreg,kawa,alekseev}.

Recently, these ideas were generalized to also treat 
$N$-terminal starlike LL/nanotube networks ($N>2$) 
\cite{chen02}.  Such a transport theory is then able to make
concrete predictions for experiments, e.g.~on carbon nanotube Y junctions.
First steps to the production and experimental analysis of SWNT
networks are under way. Template-based chemical vapor deposition
\cite{expy} and electron beam welding methods \cite{terrones} allow to
fabricate and contact multi-terminal nanotube junctions. Recently, 
an intrinsically nonlinear  current-voltage characteristics of multi-walled
nanotube Y junctions
has been reported \cite{exp2}. 
Furthermore, four-terminal devices have been realized by several
groups using two crossed NTs \cite{crossed,kims}.
Such nanotube-based (or related)
 setups are of growing interest to many theorists working
in the field of interacting mesoscopic systems. 
To mention some recent work, in Ref.~\cite{nayak} 
a detailed renormalization group analysis of a similar model
as ours was performed.
Other authors used perturbation theory
in the hopping \cite{safi2,crepieux} or in the interaction strength \cite{lal}.
Furthermore, the conductance of a junction of three
Luttinger liquids enclosing a magnetic flux was investigated \cite{chamo03}. 
The cited activity underlines the need for a generalization of the
LB approach to correlated systems.  With this work we hope to 
contribute a significant step towards this aim.  Some of our
results have been published before \cite{chen02}, but most
of the material is new.

The paper is organized as follows. In Sec.~\ref{theo} we describe our
model for the interacting multi-terminal nanotube device,
and outline its general solution in the symmetric case. 
As the step from $N=3$ to $N>3$ is straightforward \cite{chen02}, 
we concentrate on nanotube Y junctions ($N=3$) in the following sections. 
In Sec.~\ref{yjunction} the
nonlinear conductance matrix of a symmetric Y junction is discussed.
The case of asymmetric junctions is then studied in 
Sec.~\ref{asyj}.  Finally, we conclude and point
out some open questions in Sec.~\ref{conclu}.

\section{General model}
\label{theo}

\begin{figure}
\begin{center}
\epsfysize=6cm
\epsffile{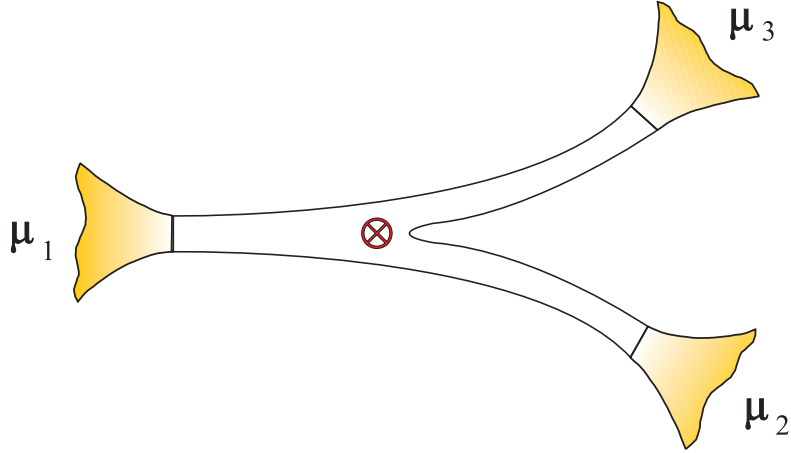}  
\end{center}
\caption{\label{fig1} Basic setup: $N$ nanotubes form a starlike
junction (shown for $N=3$). 
Each nanotube is supposed to be adiabatically contacted,
with chemical potential $\mu_i$ of the respective reservoir.
The cross indicates an additional impurity.
}
\end{figure}

We shall first discuss the simplest case of a spinless single-channel
Luttinger liquid.  In fact, in the first part (Sec.~\ref{noint}),
we even ignore the interactions and briefly review some aspects
of LB theory
relevant to our purposes.  The case of a symmetric junction is
then treated by various boundary conditions for the
interacting system in Sec.~\ref{withint},
and the extensions necessary to treat nanotubes are discussed in 
Sec.~\ref{sec23}.
These include the electronic spin and an additional
flavor degree of freedom due to the presence of two K points
in the first Brillouin zone \cite{egger97}.
The basic setup studied here is shown schematically in Fig.~\ref{fig1}.

\subsection{Fermi liquid case}
\label{noint}

When one ignores the Coulomb interaction among electrons,
transport in a starlike setup of $N$  
nanotubes can be described by the scattering matrix
approach, employing  a unitary $S$ matrix with entries
$s_{ij}$ determining the transmission coefficients
$T_{ij}=|s_{ij}|^2$ of the system. The current through nanotube $i$ 
may we written as 
\[
I_i = \frac{e^2}{h} \int_{-\infty}^\infty \ dE \Bigl\{
[1-T_{ii}(E)]f(E-\mu_i) -\sum_{j\neq i} T_{ij}(E) f(E-\mu_j) \Bigr\}  ,
\]
where $f(E)$ is the Fermi function
(we consider only a single mode here, while a 
true SWNT has four).
Unitarity of the $S$ matrix implies 
(1) the Kirchhoff rule (charge conservation),
$\sum_i T_{ij} = 1$,
and (2) gauge invariance under a
uniform potential shift in all the reservoirs,
$\sum_j T_{ij} = 1$.
When all transmission
coefficients are energy-independent,
the nonlinear conductance matrix normalized to $e^2/h$ reads
\begin{equation} \label{nonling}
G_{ij} = \frac{h}{e} \frac{\partial I_i}{\partial \mu_j} 
= \delta_{ij}-T_{ij}.
\end{equation}
Without applied magnetic field, time-reversal symmetry implies
$T_{ij}=T_{ji}$. 
For concrete calculations of the $S$ matrix for
nanotube Y junctions, we refer to Refs.~\cite{damato,ami,menon}. 
The $N\times N$ matrix $S$ relates the outgoing and incoming scattering states
at the node,
\begin{equation}\label{smat}
\Psi_L(0)= S \Psi_R(0) .
\end{equation}
To describe the $N$ (for now semi-infinite) 
tubes, we use $\Psi(x)=(\psi_1,\ldots,\psi_N) = \Psi_L+\Psi_R$ with $x\leq 
0$, which contains the outgoing (left-moving, L) components as
well as the incoming (right-moving, R) states, with  the junction 
at $x=0$.  Unfortunately, the boundary condition
(\ref{smat}) is exceedingly difficult to handle for correlated electrons.

To make progress, we shall first consider only symmetric junctions.
A wide class of such junctions can be parameterized  by a
symmetric $S$ matrix of the form
\begin{equation}\label{smatspec}
S = \left( \begin{array}{cccc} \frac{2}{N+i\lambda}-1 & \frac{2}{N+i\lambda} 
& \cdots & \frac{2}{N+i\lambda} \\ \frac{2}{N+i\lambda} & \frac{2}{N+i\lambda}-1 & 
\cdots & \frac{2}{N+i\lambda} \\ \cdots & \cdots & \cdots & \cdots \\ \frac{2}{N+i\lambda} 
& \frac{2}{N+i\lambda} & \cdots &
\frac{2}{N+i\lambda}- 1  \end{array} \right) ,
\end{equation}
where $\lambda\geq 0$ serves as arbitrary parameter.
This is the most general $S$ matrix that implies wavefunction
matching at the node \cite{texier},
\begin{equation} \label{wfm}
\psi_1(0)= \dots = \psi_N(0),
\end{equation}
for the components of $\Psi = \Psi_L + \Psi_R$. The 
$S$ matrix (\ref{smatspec}) leads to 
\begin{equation} \label{lbtransmi}
T_{ii}= \frac{(N-2)^2+\lambda^2}{N^2+\lambda^2} \; , \;\;  T_{i\neq
j}= \frac{4}{N^2+\lambda^2} ,
\end{equation}
which, for $\lambda=0$, describes a maximally transmitting
 symmetric junction,
$T_{ii}=(N-2)^2/N^2$ and $T_{i\neq j}= 4/N^2$.  On the other hand,
for $\lambda\to \infty$, 
a very bad junction forms, $T_{ii}=1$ and $T_{i\neq j}=0$.
For arbitrary $\lambda$,  the conductance matrix then 
follows from Eq.~(\ref{nonling}).
It is worth noting that no symmetric starlike junction 
can act as an ideal beam splitter, since 
backscattering at the node is unavoidable. 
The only possibility to have an ideal beam 
splitter, $T_{ii}=0$, arises in asymmetric junctions 
\cite{ami}. 

Since phases in $\psi_i(0)$ can be gauged away, a density matching condition 
arises for any symmetric junction described by Eq.~(\ref{smatspec}),
\begin{equation} \label{densmatch}
\rho_1(0) = \dots = \rho_N(0) ,
\end{equation}
where $\rho_i(0)$ denotes the total electronic density in tube
$i$ at the node.   
 The condition (\ref{densmatch}) is crucial 
in our approach for interacting systems, as it avoids the
explicit wavefunction matching.
The density in Eq.~(\ref{densmatch}) refers to the sum of the slow
(long wavelength) part, $\rho^0_i(x)=
\Psi_{R,i}^\dagger \Psi_{R,i}^{} + \Psi_{L,i}^\dagger \Psi_{L,i}^{}$,
and the Friedel oscillation,
$\delta \rho_i (x) = \Psi_{R,i}^\dagger \Psi_{L,i}^{} + \Psi_{L,i}^\dagger
\Psi_{R,i}^{}$.  Since the parameter $\lambda$ will affect the 
Friedel oscillation, there is a hidden dependence on $\lambda$ not
directly visible in Eq.~(\ref{densmatch}).

\subsection{Luttinger liquid case}
\label{withint}

Next we include the bulk interaction in the tubes, but still restrict
ourselves to a symmetric junction.  We assume that all nanotubes
can be characterized by the same interaction strength, and for now
still consider just the spinless single-channel version.  
Following Ref.~\cite{egger97}, we may neglect backscattering and
therefore obtain a Luttinger liquid, where interactions are
encoded in the standard Luttinger parameter $g\leq 1$, with
$g=1$ referring to the noninteracting limit.
Under bosonization \cite{gogol}, the
Hamiltonian is (we put $\hbar= 1$)
\begin{equation} \label{llham}
H = \frac{v_F}{2} \sum_{i=1}^N \int_{-L}^0 \ dx \ \left[ \Pi_i^2+\frac{1}{g^2}
\left(\partial_x \theta_i \right)^2 \right] \; ,
\end{equation}
where $\Pi_i(x)$ is the canonical momentum to the standard boson field
$\theta_i(x)$ for tube $i$, and $v_F$ is the Fermi velocity. 
Similar to the two-terminal case, adiabatically
coupled external voltage reservoirs (at $x\approx -L$) 
lead to radiative boundary 
conditions for the long-wavelength part of the chiral
electron densities \cite{egger96},
\begin{equation}\label{bc}
 (g^{-2}+1)\rho^0_{i,R}(-L)  + 
 (g^{-2}-1)\rho^0_{i,L}(-L)  = \mu_i/\pi v_F .
\end{equation}
These conditions only depend on the injected
currents,  which are independent of any
backscattering happening at the node or within the nanotube. 
To fulfill Eq.~(\ref{bc}),
the chiral densities must combine to give
\begin{eqnarray} \label{ans1}
I_i &=& ev_F(\rho^0_{i,R}-\rho^0_{i,L})=
 \frac{e^2}{2\pi} \left(U_i-\sum_j \tilde{T}_{ij} U_j \right), \\ \label{ans2}
\rho^0_i &=& \rho^0_{i,R}+ \rho^0_{i,L} =
 \frac{e g^2}{2\pi v_F} \left (U_i + \sum_j \tilde{T}_{ij} U_j \right) ,
\end{eqnarray}
which defines effective transmission matrix elements
$\tilde{T}_{ij}(g, \{\mu_k\}, k_B T)$.
The matrix $\tilde{T}$ should not be confused
with the LB transmission matrix $T$ of the noninteracting system. 
Albeit $\tilde{T}= T$  for $g=  1$,  the
effective transmission matrix $\tilde{T}$ cannot be
interpreted as single-electron transmission matrix for $g<1$.
The applied voltages in Eqs.~(\ref{ans1}) and (\ref{ans2}) are defined as
\begin{equation} \label{ui}
eU_i = \mu_i - \bar{\mu} \; , \;\; \bar{\mu} = \sum_i \mu_i/N  .
\end{equation}
Under this definition, gauge invariance is automatically fulfilled if the
transport properties of the system only depend on the $U_i$ and not explicitly
on the $\{\mu_k\}$.
Charge conservation (Kirchhoff's rule) can be incorporated by imposing
\begin{equation} \label{tildetij}
\sum_{i=1}^N \tilde{T}_{ij} = 1 .
\end{equation}
Finally, the density matching conditions (\ref{densmatch}) have to be obeyed. 

Due to the pinning amplitude $\delta \rho_i(0)$, i.e.~the
value of the Friedel oscillation at the node,
 $\rho_i(0)$ will deviate from
$\rho^0_i$ in Eq.~(\ref{ans2}). 
To compute the pinning amplitude, we first note
that the Friedel oscillation
$\delta\rho_i(x)$ in tube $i$ arises due to interference of the incoming
right-movers and the left-movers that are backscattered at the node.
Left-movers transmitted from the other $N-1$ tubes
into tube $i$ cannot interfere with the incoming
 right-movers and will therefore not
contribute to $\delta \rho_i(x)$. 
To verify this argument explicitly, it suffices to analyze the
corresponding two-terminal case with one impurity 
at $x=0$.  Without loss of generality, let us take $k$-independent
bare transmission amplitude $t$ for the impurity.
Taking $k>0$, there is a right-scatterer eigenstate $\phi_k(x)$,
\[
\phi_k(x) = \frac{1}{\sqrt{2\pi}}
 \left \{ \begin{array}{cc}   e^{ikx}+r e^{-ikx} & (x<0)\\
t e^{ikx} & (x>0) \end{array}\right. ,
\]
where $|t|^2+|r|^2=1$.  Similarly, the left-scatterer is
 $\phi_{-k}(x)= \phi_k(-x)$.  We now show
that there is no contribution to $\delta \rho(x)$ 
from interfering right- and left-scatterers. While this
statement is immediately seen for noninteracting electrons,
it can also be confirmed for $g<1$.  To that
purpose, let us expand the electron operator in terms of the
right- and left-scatterers instead of the standard R/L movers,
\begin{equation}\label{rls}
\Psi(x) \sim \sum_{k>0}  a_k \phi_k(x) + b_k \phi_{-k}(x).
\end{equation}
Of course, the kinetic part of the Hamiltonian decouples in the
Fermi operators $a_k$ and $b_k$.
Moreover,  when inserting Eq.~(\ref{rls}) into the interaction
Hamiltonian, no terms containing only one $a_k$ or $b_k$ 
can appear because the right- and left-scattering states
$\phi_k(x)$ and $\phi_{-k}(x)$ are
orthonormal.  As a consequence, in the expectation value
of the density operator, there can be no terms involving
$\langle a^\dagger_k b_k^{} \rangle$ or $\langle b^\dagger_k a_k^{}
\rangle$.  Hence there is no interference between scattering states
originating in different reservoirs, a finding that directly 
translates to the $N$-terminal setup under study here.

As a consequence of the above discussion, $\delta\rho_i(0)$ must be
identical to the pinning amplitude in a related two-terminal
setup defined by an impurity whose bare reflection probability is
 determined by $T_{ii}$. 
Using this mapping, we can thus compute the pinning amplitude from
\begin{equation} \label{rvrela}
\langle \delta \rho_i(0) \rangle = - g^2 e V_i/\pi v_F ,
\end{equation}
where the four-terminal voltage $V_i = V_i(U_i)$ has to be determined
through a self-consistency relation, see below.
The validity of Eq.~(\ref{rvrela}) can be seen as follows. 
The long-wavelength density fluctuation in the equivalent
two-terminal setup, due to an applied
voltage $U_i$ as given by Eq.~(\ref{ui}),
is \cite{egger96}
\[
\langle\rho^0(x)\rangle = - g^2 eV_i \ {\rm sgn}(x) / 2\pi v_F.
\]
With respect to a density corresponding to the chemical potential $\bar{\mu}$,
taking $x<0$, we thus have $\rho^0_i=  g^2 eV_i(U_i)/\pi v_F$.  
For the two-terminal problem, antisymmetry 
of $\langle\rho(x)\rangle$ under $U\to -U$ implies that the net
change in density due to $U$ 
is zero at $x=0^-$.  Therefore $\rho^0_i$
must be cancelled by the pinning amplitude, 
see Eq.~(\ref{rvrela}).  Note that this
reasoning does not assume the validity of the density matching 
conditions (\ref{densmatch}).

For $g=1/2$, the self-consistency relation determining $V(U)$ 
can be obtained using refermionization, and takes a rather simple form.
Results for this special interaction strength
have been discussed in Ref.~\cite{chen02}.  
Using integrability concepts, however,  the respective
self-consistency relation
can also be derived and solved for arbitrary $g$
\cite{egger00}. 
Let us write down the necessary equations for
$g=1/p$ (with integer $p$), keeping in mind that
arbitrary values of $g$ can also be treated.
The bare reflection probability of the impurity determines the
 energy scale
$k_B T_B\propto \exp\theta_B$, see below, and the elementary
excitations scattered one-by-one at the impurity
 are kinks, antikinks, and $p-2$ neutral breathers.
The two-terminal current $I$ then follows as an integral
over rapidities $\theta$, 
\begin{equation}\label{self1}
I= \int |T_{++}|^2  (\sigma_+ - \sigma_-) d\theta,
\end{equation}
where $|T_{++}|^2 = 1/\{1+ \exp[-2(p-1)(\theta-\theta_B)]\}$.
The occupation probabilities are $\sigma_\pm=n f_\pm$ for the kinks/antikinks,
where $2\pi n= d\epsilon/d\theta$ with the eigenenergy $\epsilon$
of the kink/antikink. Explicit expressions for the 
eigenenergies can be found in Ref.~\cite{egger00}. 
Furthermore, the filling fractions read 
$f_\pm(\theta)=1/(1+\exp[(\epsilon(\theta)\mp W/2)/T])$,
where $\pm W/2$ is an effective chemical potential for the kink/antikink.  The
parameter $W$ has to be determined self-consistently through
\begin{equation}\label{self2}
\int \left( |T_{++}|^2+ p |T_{+-}|^2 \right) (\sigma_+-\sigma_-)d \theta
= g e U/2\pi v_F ,
\end{equation} 
where $U$ is the applied voltage and
 $|T_{+-}|^2=1-|T_{++}|^2$.  Since 
$(e^2/2\pi)(U-W) =  (1-p) I$, see Ref.~\cite{egger00},
the solution of  Eqs.~(\ref{self1}) and (\ref{self2}) then yields 
$W=W(g, k_B T,U)$, and finally the
four-terminal voltage follows from $V=W-2\pi I/ge^2$. 

In these equations, $T_B$ appears as effective impurity
strength of the related two-terminal problem.  With
$T_{ii}$ given in Eq.~(\ref{lbtransmi}),
an effective dimensionless impurity strength $v_0$ 
can be determined by examining the
noninteracting limit, see below.
Given $v_0$, the scale $T_B$ then follows as
\begin{equation}\label{tb}
k_B T_B/D  =  c_g v_0^{ 1/(1-g)}, 
\end{equation}
where $D$ is the electronic bandwidth and $c_g$ a numerical
prefactor of order unity whose precise value is given in
Ref.~\cite{egger00}.
Using this correspondence, the complete transport problem for
a symmetric junction described by the bare $S$ matrix (\ref{smatspec})
can be solved explicitely.  Using Eq.~(\ref{rvrela}),  
the density matching conditions (\ref{densmatch}) read
$(1+\tilde{T}_{jj})U_j-2V_j = \tilde{T}_{kj}U_j$ 
for all pairs $k\neq j$.  Exploiting Kirchhoff's rule
(\ref{tildetij}), this set of equations admits only the solution
\begin{equation} \label{Tiig} 
 \tilde{T}_{ii}  = \frac{2-N}{N}+ \frac{2(N-1)V_i}{N U_i} \;, \quad 
 \tilde{T}_{k\neq i}  = \frac{2}{N}- \frac{2V_i}{N U_i}. 
\end{equation} 
Evidently, the $\tilde{T}_{ij}$ depend only on the applied voltages $U_i$ but
not separately on the chemical potentials $\mu_i$. Thus gauge invariance is
automatically fulfilled.  The conductance matrix then follows 
immediately from Eq.~(\ref{ans1}).  
In Sec.~\ref{yjunction}, we will discuss this solution in some
detail for the case $N=3$ (nanotube Y junction).  It  then only remains
to fix the impurity scale $v_0$ employed in determining $V_i(U_i)$.
 
As the impurity strength parameter $v_0$ 
can be obtained in the noninteracting limit,
a convenient way to fix $v_0$ is as follows.
For $g=1$, as pointed out above,  $\tilde{T}_{ii}$
should simply reproduce the
LB transmission coefficient $T_{ii}$ specified in Eq.~(\ref{lbtransmi}).
In addition, in the noninteracting limit, the ratio $V_i/U_i$ 
is just the bare reflection coefficient $R$ of the impurity in the
two-terminal problem \cite{egger96}.  The latter
can be expressed in terms of $v_0$ according to
\[
R =  \frac{v_0^2}{(1+v^2_0/4)^2} ,
\]
where $v_0=2$ corresponds to the unitary limit; larger
values for $v_0$ lead to unphysical results. Combining above equations
then leads to
\begin{equation}\label{v0def}
v_0(\lambda,N) = 2(\sqrt{N^2+\lambda^2} - \sqrt{2N})/\sqrt{N(N-2)+\lambda^2}. 
\end{equation}
For a very bad junction, $\lambda\to \infty$, we are indeed in the
unitary limit, $v_0=2$.  For a perfect junction with $\lambda=0$, however,
$v_0=2(\sqrt{N} - \sqrt{2})/\sqrt{N-2}$.  In particular,
for $N=3$, this value is $v_0\approx 0.63567$.

\subsection{Extensions}
\label{sec23}

Real SWNTs are characterized by four channels, resulting from
the electron spin and the additional flavor degree of freedom.
Including these degrees of freedoms as additional indices $(\sigma,\alpha)$,
Eq.~(\ref{smat}) now involves a $4N \times 4N$ dimensional $S$ matrix,
and the electron state $\Psi=\Psi_R+\Psi_L$ has $4N$ entries
$\psi_{i\sigma\alpha}(x)$.   Assuming that at the node neither
flavor mixing nor spin-flip processes are present,  
the previous analysis can largely be carried over. 
Since the $S$ matrix factorizes, the Kirchhoff rule as well
as the density matching conditions hold
separately for each component $(\sigma\alpha)$.
The Hamiltonian (\ref{llham}) then corresponds to a four-channel
boson model \cite{egger97} in terms of symmetric/antisymmetric
charge/spin fields. The interaction parameter in the
symmetric charge mode is then $g_{c+}<1$,
while the neutral modes are essentially at the noninteracting
value.  The radiative boundary conditions (\ref{bc})
are only imposed in the symmetric charge sector,
and our above discussion then directly leads to a very similar
expression as in Eq.~(\ref{ans1}).
However, when computing the effective transmission matrix $\tilde{T}$,
 the four-terminal voltage parameter $V_i$
is  now nontrivially affected by the additional degrees of freedom.
 This is due to the fact that backscattering at the node 
couples the four boson modes diagonalizing the Hamiltonian.
At the moment, we are not aware of an exact solution to this problem.
 Nevertheless, two remarks can be made even in this
situation.  First, the prefactor in the
current (\ref{ans1}) is changed from $e^2/h$ to $4e^2/h$, 
reflecting the presence of four channels.
Second,   we expect that the typical
power law exponents to lowest order are modified
according to $g \to (g_{c+}+3)/4$. Therefore, the 
qualitative physics seems to be captured already by the
one-channel version discussed further on.  
However, we wish to stress
that the finite value for $v_0$ obtained under the mapping does
not seem to allow for simple perturbative access to the full 
four-channel problem.

\section{Conductance of a symmetric Y junction}
\label{yjunction}

\begin{figure}
\vspace{0.5cm}
\begin{center}
\epsfysize=9cm
\epsffile{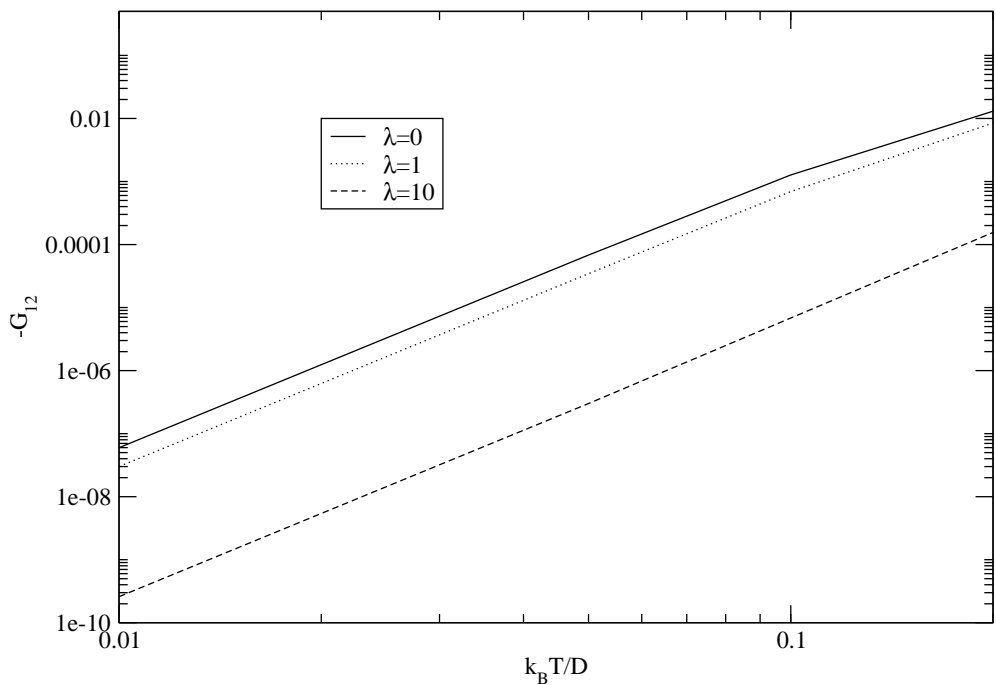}  
\end{center}
\caption{\label{fig2} Temperature dependence of the
linear conductance coefficient $G_{12}$ for $g=1/3$
  on a double-logarithmic scale, for
several different values of the parameter $\lambda$ in Eq.~(\ref{smatspec}).
}
\end{figure}

\begin{figure}
\vspace{0.5cm}
\begin{center}
\epsfysize=9cm
\epsffile{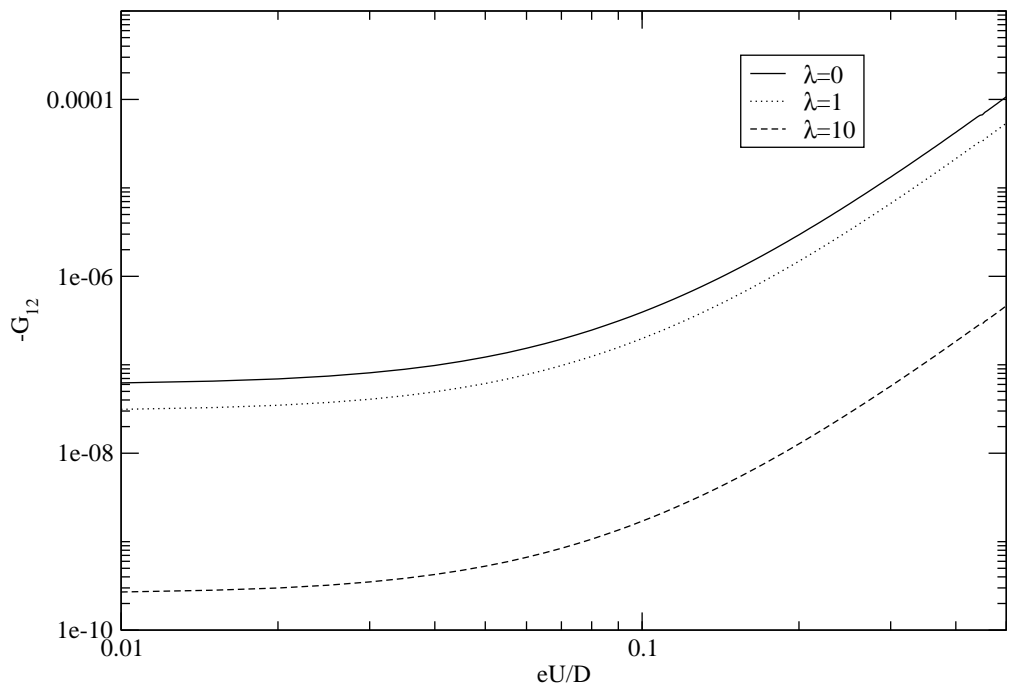}  
\end{center}
\caption{\label{fig3} Voltage dependence of the nonlinear
conductance $G_{12}$ (double-logarithmic scale) at $k_B T/D=0.01$,
for several values of $\lambda$.
}
\end{figure}

In this section we discuss the conductance defined by
$G_{ij}=(h/e)\partial I_i/\partial \mu_j$
 for the case of a symmetric Y junction ($N=3$), based on
the solution of the general problem in Sec.~\ref{withint}. Using 
Eq.~(\ref{Tiig}) gives
the nonlinear conductance matrix in the form
\begin{eqnarray}
G_{ii} &=&  \frac{8}{9} \left(1-\frac{\partial  V_{i}} {\partial U_{i}} \right)
+ \frac{2}{9} \sum_{j\neq i} \left(1 -  \frac{\partial  V_{j}} {\partial
U_{j}} \right), \label{giig}
 \\
G_{j\neq i} &=&   \frac{4}{9}  \left( \frac{\partial  V_{i}} {\partial U_i} - 1 \right)
                + \frac{4}{9}  \left( \frac{\partial  V_{j}} {\partial U_j} - 1 \right)
                -  \frac{2}{9}  \left( \frac{\partial V_{k}} {\partial U_k} -
1 \right), \nonumber
\end{eqnarray}
where $k\neq i \neq j$ in the second equation.
Eq.~(\ref{giig}) holds for any interaction strength $g$. 
To obtain explicit results, one only needs to compute $V(U)$ for
the corresponding two-terminal problem.  We shall show explicit results
now for several values of $\lambda$,
the special interaction parameter $g=1/3$, and applied 
voltages $\mu_1=E_F+ eU$, $\mu_2=\mu_3=E_F$.
In Fig.~\ref{fig2}, the linear conductance is shown,
and in Fig.~\ref{fig3} the nonlinear conductance is plotted
for $k_B T/D=0.01$.

As illustrated in Figs.~\ref{fig2} and \ref{fig3} for $G_{12}$ at
$g=1/3$, at low energies the system always flows to 
a  fixed point corresponding to disconnected nanotubes,
\begin{equation} \label{fixed}
G_{ij} = 0.
\end{equation}
To the best of our knowledge, 
all theoretical work on this problem
\cite{chen02,nayak,safi2,crepieux,lal,chamo03} 
agrees that for symmetric junctions, Eq.~(\ref{fixed}) represents
the only stable fixed point in the case of repulsive
electron-electron interaction. 
This fixed point represents an isolated node weakly coupled to
$N$ broken-up quantum wires.  
It is then of interest to study the energy dependence of the $G_{ij}$
at low energies, where a typical LL $g$-dependent power law
behavior arises, see also Ref.~\cite{lal}. 
The leading corrections to Eq.~(\ref{fixed})
can therefore be formulated in terms of the well-known end tunneling
density of states, with associated power law exponents 
given in textbooks \cite{gogol}.

\section{Conductance of an asymmetric Y junction}
\label{asyj}

Next, we construct more general $S$ matrices starting from
the solution for the symmetric situation discussed in the two previous
sections, focussing for clarity on the case $N=3$.
The basic idea to make the system asymmetric is to add
impurities of dimensionless strength $v_i$ 
to tube $i$. To avoid the appearance of 
spurious resonances, these are taken sufficiently close to the
node, at, say, $x \approx -1/k_F$.  This modelling of an 
asymmetric junction still allows for $g<1$, e.g.~by
using perturbation
theory in the $v_i$ around the above solution for a
symmetric junction ($v_i=0$).
The prototypical 
situation for exactly one impurity, $v_1 > 0$ but $v_2=v_3=0$, 
is  illustrated in Fig.~\ref{fig1}. 
Since the impurities are displaced away from $x=0$, the junction
itself is characterized by the symmetric $S$ matrix (\ref{smatspec}),
and therefore the density matching conditions (\ref{densmatch}) can
still be used.  Nevertheless, effectively this procedure allows
to generate asymmetric $S$ matrices.

The free real-time
 action then reads in the framework of bosonization \cite{gogol}
\begin{equation} \label{freeac}
S_0[\theta_i]  = \frac{v_F}{2} \sum_{i=1}^3 \int_{-L}^0 dx \int dt \
 \left[ \frac{1}{v_F^2 } \left(\partial_t
 \theta_i \right)^2 - \frac{1}{g^2} \left(\partial_x \theta_i
\right)^2 \right]  ,
\end{equation}
where the bosonic phase field $\theta_i(x,t)$ is decomposed according to
$\theta_i = \theta_i^h + \theta_i^p$. Here, $\theta_i^h$ obeys  homogeneous
($U_i=0$) boundary conditions (\ref{bc}), and $\theta_i^p$ is
a particular solution to the equations of motion
obeying Eq.~(\ref{bc}). 
Written in terms of the boson fields, one has
$I^0_i(x,t) = (e/\sqrt{\pi}) \partial_t \langle \theta_i (x,t)
\rangle$ and $\rho^0_i(x,t) = (1/\sqrt{\pi}) \partial_x \langle
\theta_i (x,t) \rangle$, and one checks easily that
the equations of motion as well as  Eq.~(\ref{bc}) are satisfied by
\begin{equation} \label{parti}
\theta_i^p(x,t) = \sqrt{\pi}I^0_i t/e + \sqrt{\pi} \rho^0_i x  ,
\end{equation}
where $I^0_i$ and $\rho^0_i$ are given by Eqs.~(\ref{ans1}) and (\ref{ans2}),
respectively. Here we assume that applied voltages are chosen such that
 $\bar{\mu}=0$.

Adding an impurity of strength $\Lambda_1= D v_1/\pi$ (with bandwidth $D$)
into tube 1 then yields the action contribution
\begin{equation} \label{si}
S_I = \Lambda_1 \int dt \cos [\sqrt{4\pi} \theta_1^h (0,t) +
2\pi I^0_1 t/e].
\end{equation}
>From now on, $L \rightarrow \infty$ to allow for analytical
progress. Inserting Eq.~(\ref{parti}) into Eq.~(\ref{freeac}), we obtain 
\[
S_0[\theta_i] = S_0 [\theta^h_i] - \sqrt{\pi}v_F \sum_{i=1}^3
(\rho_i(0)/g^2) \int dt \  \theta_i^h(0,t)  .
\]
The second term can be dropped because of 
Eq.~(\ref{densmatch}) and  Kirchhoff's rule, which implies that
$\sum_i \theta_i^h(0,t)$ equals an arbitrary constant (which 
can be put to zero).
The current $I_i$ is then given by
\begin{equation} \label{iias}
I_i = \frac{e^2}{2\pi} \left(U_i-\sum_j \tilde{T}_{ij} U_j \right) +
\frac{e}{\sqrt{\pi}} \langle \partial_t \theta_i^h (0,t) \rangle  .
\end{equation}

\subsection{Weak asymmetry: Perturbation theory}
\label{sss}

The impurity strength $\Lambda_1$ measures the degree of asymmetry
of the junction. First, we treat the case of weak asymmetry under a
perturbative calculation in $\Lambda_1$, using the Keldysh technique.
In this subsection, we write $\theta^h \rightarrow
\theta$ and restrict ourselves to $T=0$.
Then  the current (\ref{iias}) follows from
\begin{equation} \label{expval}
\langle \partial_t \theta_i (0,t) \rangle = \frac{1}{Z} 
\sum_{i=1}^3 \int {\cal D} \theta_i e^{iS[\theta_i]} 
 \delta \left( \sum_i\theta_i(0,t) \right)
\partial_t \theta_i (0,t) ,
\end{equation}
where $S[\theta_i] = S_0[\theta_i] + S_I$. The
$\delta$-function reflecting Kirchhoff's rule is enforced using
a Lagrange multiplier field $\gamma(t)$, yielding the additional
action contribution $S_\gamma = \int dt \ \gamma(t) \sum_i \theta_i(0,t) 
$.  Switching to Fourier space, the
action $S[\theta_i,\gamma]=S_0[\theta_i]+S_\gamma$ reads
\[
S[\theta_i,\gamma] = \int_{-\infty}^\infty
 \frac{dk}{2\pi} \frac{d\omega}{2\pi}
\sum_{i=1}^3 \left\{ \frac{v_F}{4} \left( \frac{\omega^2}{v_F^2} -
\frac{k^2}{g^2} \right) | \theta_i(k,\omega)|^2 + \gamma(-\omega)
\theta_i(k,\omega)  \right\} .
\]
When switching to $k$-space, a principal value 
contribution has been dropped, which 
is justified in the low-energy regime of interest here \cite{bjoern}.
We then introduce rotated phase fields,
$\theta_\pm=\theta_2 \pm \theta_3$,
with conjugate momenta $\Pi_\pm =
(\Pi_2 \pm \Pi_3)/2$, and subsequently integrate out the fields
$\theta_1$ and $\gamma$ by standard Gaussian integration.
This procedure leads to the (impurity-free) effective action
 $S_0[\theta_+] + S_0[\theta_-]$, where
\begin{equation} \label{feffaction}
S_0[\theta_\pm] = \frac{v_\pm}{2} \int \frac{d\omega}{2\pi} \frac{dk}{2\pi}
\left( \frac{\omega^2}{v_\pm^2} - \frac{k^2}{g_\pm^2} \right) \left|
\theta_\pm (k,\omega) \right|^2,
\end{equation} 
with $v_+=4v_F/3, g_+=4g/3, v_-=4v_F,$ and $g_-=4g$. 
Fourier transformation back to real space gives  
\begin{equation} \label{effaction}
S_0[\theta_\pm]=\frac{v_\pm}{2} \int_{-\infty}^\infty
dx dt \left[ \frac{1}{v_{\pm}^2} (\partial_t \theta_\pm)^2
- \frac{1}{g_{\pm}^2} (\partial_x\theta_\pm)^2\right] ,
\end{equation}
where the rotated boson fields $\theta_\pm(x)$ live on the
full line.  As we shall see below, only a certain linear combination 
of the right- and left-moving components of $\theta_+$ couples
to the impurity, and hence the extension to the positive half-line
($x>0$) does not introduce
spurious effects while simplifying the subsequent analysis 
considerably.

Expectation values of the form (\ref{expval})
are then calculated by first evaluating 
\begin{equation} \label{expval2}
\langle \partial_t \theta_\pm (t) \rangle = \frac{1}{Z} \int {\cal D}
\theta_+ {\cal D}\theta_- e^{i(S[\theta_+]+S_0[\theta_-])} \partial_t
\theta_\pm (t) ,
\end{equation}
where the impurity in tube 1 acts only in the $+$ sector,
\begin{equation} \label{stp}
S[\theta_+] = S_0[\theta_+] + \Lambda_1 \int dt \cos [ - \sqrt{4\pi}
\theta_+ (0,t) + 2\pi I_1^0 t/e ].
\end{equation}
Since $\langle \partial_t \theta_- (t) \rangle = 0$,
we are left with the relations
$\langle \partial_t \theta_1  \rangle = -
 \langle \partial_t \theta_+ \rangle$ and 
$\langle \partial_t \theta_{2,3}  \rangle = 
\langle \partial_t \theta_+  \rangle/2.$
Equation (\ref{expval2}) can now be evaluated perturbatively in $\Lambda_1$,
and Eq.~(\ref{iias}) then gives  \cite{bjoern}
\begin{equation}\label{solone}
I_1 = I^0_1 - \delta I_{\Lambda_1}, \quad 
I_{2,3}= I_{2,3}^0 + \delta I_{\Lambda_1}/2,
\end{equation}
where $I_i^0$ denotes the currents
computed in Sec.~\ref{withint} in the absence of the impurity,
and the lowest-order perturbative correction due to the impurity is
\begin{equation} \label{deli}
\delta I_{\Lambda_1} = e g_+ (\Lambda_1^2/D) \sin(2 \pi g_+) \Gamma(1-2 g_+) 
{\mbox {sgn}} (I^0_1) ( 2\pi |I^0_1|/e D)^{2g_+ -1} .
\end{equation}

To treat more general $S$ matrices, one can add a
second (weak) impurity $\Lambda_2$, say, into tube 2. 
Then, within order $\Lambda_1^2$, again 
correction (\ref{deli}) arises, the correction of order $\Lambda_1 \Lambda_2$
vanishes, and corrections of order $\Lambda_2^2$ may be
calculated in the same way as above. This gives
\begin{eqnarray} \label{deli2}
\delta I_{\Lambda_2}^\pm &=& \frac{e}{\sqrt{\pi}}  \langle
\partial_t \theta_\pm (0,t) \rangle\nonumber \\
&=& - e g_\pm (\Lambda_2^2/2 D) \sin
(2 \pi g_+ ) \Gamma (1-2g_+) {\mbox {sgn}} (I^0_2)
(2\pi |I^0_2|/e D)^{2 g_+ -1}  ,
\end{eqnarray} 
where we replaced most of the $g_-$ that actually appear in Eq.~(\ref{deli2})
by $g_+$, utilizing $g_+ + g_- = 4g_+$,
and the currents $I_i$ are 
\begin{eqnarray*} 
I_1 &=& I^0_1 - \delta I_{\Lambda_1} - \delta I_{\Lambda_2}^+ , \\ 
I_{2,3} &=& I^0_{2,3} +  (\delta I_{\Lambda_1} + \delta
I_{\Lambda_2}^+ \pm \delta I_{\Lambda_2}^- )/2 .
\end{eqnarray*}
Evidently, a wide class of asymmetric Y junctions is accessible by this
construction method even for interacting SWNTs.

\subsection{Arbitrary asymmetry: Refermionization}

In the case of just one impurity ($\Lambda_1$), it is possible
to solve the full transport problem in a nonperturbative manner
at the special interaction strength $g=3/8$. This point
corresponds to the Toulouse limit $g_+=1/2$ in the rotated picture 
used in Sec.~\ref{sss}, and allows to use refermionization \cite{gogol}.
In fact, one can solve this transport problem
for arbitrary $g$ using integrability
methods, see Ref.~\cite{egger00}, but for clarity, in this section
we only discuss the simplest non-trivial case, $g=3/8$, which
admits the use of standard techniques.
We only have to study the $\theta_+$ sector, as the
$\theta_-$ field decouples from the impurity and
can be ignored.  

To sketch the $g=3/8$ solution, let 
us first define new chiral boson fields ($\phi_+(x)$ is
the conjugate field, $\Pi_+=\partial_x\phi_+$)
\[
\varphi_{R/L}(x) = \sqrt{\pi} (
\sqrt{g_+} \phi_+(x) \pm  \theta_+(x) /\sqrt{g_+} \ ),
\]
which obey the algebra
$[\varphi_{p}(x),\varphi_{p'}(x')]=-i\pi \delta_{pp'} {\mbox{sgn}} 
(x-x')$.
Next we switch to linear combinations of these chiral fields,
$\varphi_{\pm} = ( \varphi_{R} \mp \varphi_{L})/\sqrt{2}$,
with associated densities
$\rho_{\pm} =  \partial_x \varphi_{\pm} /2\pi$.
Then the effective Hamiltonian reads
\begin{equation} \label{fullhami}
H = \frac{v_+}{4\pi g_+} \int_{-\infty}^\infty
 dx [ (\partial_x \varphi_+ )^2
+ (\partial_x \varphi_- )^2 ] + \Lambda_1 \cos
(\sqrt{2g_+} \varphi_+(0) ) , 
\end{equation}
where the radiative boundary conditions are
$3 \rho_+(L) - \rho_+(-L) = 2 g_+ I_1^0/ev_+$, and
$3 \rho_-(L) + \rho_-(-L)  = 0$.
Therefore, only the $\varphi_+$ boson contributes to transport.
Next we introduce effective fermion fields,
$\tilde{\psi}(x) = \sqrt{k_F/2\pi} \exp(i \varphi_+(x))$.
It is convenient to introduce yet another fermion $\psi(x)$ via
$\tilde{\psi}(x) = (c+c^\dagger) \psi(x)$,
where $c$ is an auxiliary fermion.  This standard procedure
is explained in detail in the textbook  \cite{gogol}. 
In the fermionized framework,  Eq.~(\ref{fullhami}) becomes
\begin{equation} \label{fermiham}
H = -\frac{iv_+}{g_+} \int dx \ \psi^\dagger \partial_x \psi  +
\sqrt{\frac{v_+ \Lambda_B}{2g_+}} (c+c^\dagger) \left( \psi(0)-\psi^\dagger(0)
\right), 
\end{equation}
with $\Lambda_B=\pi\Lambda_1^2/D$ determining the
degree of asymmetry of the junction. In this form, the Hamiltonian can be
diagonalized immediately, and following Refs.~\cite{gogol,egger96,egger00},
a self-consistency relation determining $\delta
I_{\Lambda_1}$ follows 
\begin{equation} \label{reffini}
\delta I_{\Lambda_1} = (e\Lambda_B/\pi)  {\rm Im} \psi \left(
\frac{1}{2} + \frac{\Lambda_B + 2\pi i 
(I^0_1 - \delta I_{\Lambda_1}/2)/e}{2\pi k_B T} \right)  ,
\end{equation}
where $\psi$ is the digamma function.
The currents $I_i$ are then given by Eq.~(\ref{solone}),
with $\delta I_{\Lambda_1}$ obtained as  the
self-consistent solution of Eq.~(\ref{reffini}).
One checks easily that the $T=0$ corrections for small $e\Lambda_B/I_1^0$
predicted by this solution coincide with our previous perturbative results,
see Eq.~(\ref{deli}) for $g_+=1/2$. 
In the opposite limit of strong asymmetry, we find the expansion
\begin{equation}
\delta I_{\Lambda_1} / I^0_1 =  1 -  (\pi^2/6) (I^0_1/e\Lambda_B)^2 +
\dots 
\end{equation}
This corresponds to the highly asymmetric case with almost no current through
tube 1 and the currents through tubes 2 and 3 given by $I_{2/3} =
I^0_{2/3} + I^0_1/2$.

Remarkably, this procedure allows to nonperturbatively
 treat the full crossover between a symmetric
Y junction and a single SWNT that is approached by the tip of another
SWNT at some point in its bulk.  The corresponding 
asymptotic power law exponents
therefore correspond to the sum of end and bulk exponents in the
tunneling density of states \cite{gogol}.
Finally, we note
 that the situation of a perfect beam splitter cannot be described by
our model. This is due to the initial symmetry of our setup without
impurities, while a beam splitter can only be realized
if transport between two of the three tubes is totally suppressed
\cite{ami}.   

\section{Conclusions}
\label{conclu}

In summary, we have proposed a generalization of Landauer-type
transport theory to $N$-terminal starlike interacting nanotubes.
Due to the formulation of three
different boundary conditions, reflecting the coupling to external leads,
charge conservation, and density matching conditions
 at the node, it has been possible to find an analytical
solution for the nonlinear conductance matrix of the 
symmetric system for arbitrary electron-electron interaction.
Based on the symmetric solution, we have constructed a wide class
 of asymmetric
transmission matrices by adding impurities close to the node in the individual
nanotubes. 
The asymmetric problem can be also be solved in a nonperturbative fashion
if just one impurity is put into a nanotube arm.  For simplicity, we
have discussed this solution for the special interaction strength $g=3/8$,
corresponding to the Toulouse limit in a folded picture.

A largely open issue is the question of noise and particle statistics in such
a  nonchiral multi-terminal Luttinger liquid network. 
For low energies, the disconnected-node fixed
point seems to hinder the observation of 
fractional quasiparticle cross-correlations,
because tunneling of electrons from one tube to the other is the first
allowed process once the applied voltage is raised. However, for intermediate
energies, which could be temperature and/or applied voltage in
experiments, the disconnected fixed point opens up and quasiparticle tunneling
between different tubes might give distinct features to the current cross
correlations and therefore allow to study quasiparticle statistics.
These questions will be studied in future work. 

\ack

We wish to thank H.~Grabert, A.~Komnik, I.~Safi, and C.~Texier for useful
discussions. Support by the DFG under the Gerhard-Hess program and under Grant
No. GR 638/19 is acknowledged.

\end{document}